\documentclass{elsart}

\usepackage{graphicx}

\usepackage{url}

\begin{document}

\begin{frontmatter}



\title{Segmentation of Retinal Blood Vessels Using Deep Learning }


\author{Ifeyinwa Linda Anene and Yongmin Li}
\address{Brunel University London,
  United Kingdom}

\begin{abstract}

The morphology of retinal blood vessels can indicate various diseases in the human body, and researchers have been working on automatic scanning and segmentation of retinal images to aid diagnosis. This project compares the performance of four neural network architectures in segmenting retinal images, using a combined dataset from different databases, namely the UNet,  DR-VNet, UNet-ResNet and UNet-VGG.


\end{abstract}


\end{frontmatter}

\section{Introduction}

The ability to see is essential for individuals to function in society, but many suffer from visual impairments due to illness or other reasons beyond their control. According to the World Health Organization, an estimated 2.2 billion people have vision impairments, with at least 1 billion of those cases being preventable or treatable. Leading causes of vision loss include cataracts, age-related macular degeneration, glaucoma, diabetic retinopathy, corneal opacity, and trachoma. These conditions can affect individuals of all ages, including premature infants who are susceptible to retinopathy of prematurity, a leading cause of childhood blindness.

Despite the prevalence of eye diseases, there are many impediments to monitoring and detecting them, including low doctor-to-patient ratios, manual examination and diagnosis that requires extensive clinical training, time-consuming analysis, and subjectivity. Therefore, continuous research and development in ophthalmology are necessary to prevent a significant global population from losing their vision.

The primary focus of this project is to investigate the performance of four convolutional neural network architectures in the task of segmenting retinal blood vessels, which can aid in the early detection and treatment of eye diseases. It is essential to address these problems to ensure that individuals can maintain their sight and live healthy and productive lives.

The rest of the paper is organised as follows. The background and previous studies are reviewed in Section 2. The four different methods investigated in this project are described in Section 3. Experiments and results are presented in Section 4 before conclusions are drawn in Section 5.

\section{Background}

The human eye plays a critical role in the body since approximately 80\% of what we learn is through sight \cite{VisionInitiativeAustralia}. To see, light rays enter the eye through the cornea, which has the ability to refract and bend light rays onto the iris, determining how much light enters the eye. The lens works like a camera lens and can change its width to focus light rays properly. The light then passes through a gel-like substance called the vitreous, which helps maintain the eye's spherical shape. Finally, the lens and cornea work together to focus light accurately on the retina, a light-sensitive membrane at the back of the eye. Once the light hits the retina, it is converted to electrical signals and sent through the connected optic nerve to the brain for interpretation \cite{NationalEyeInstitute,NationalKeratoconusFoundation,VisionInitiativeAustralia}.

Eye examinations are often required when ophthalmologists encounter patients with eye ailments or other health concerns. Equipment used in an eye exam include the autorefractor, focimeter, tonometer, perimeter (visual field scanner), optomap, fundus camera, and optical coherence tomography (OCT) \cite{Elizabeth2019}. Deep learning mostly utilizes images produced from a fundus or OCT camera, with this dissertation utilizing images from a fundus camera.

The fundus camera takes photographs of the retina and is useful for eye care officials in diagnosing and monitoring the progression of many eye diseases and other health conditions. Morphological attributes of retinal blood vessels, such as length, width, tortuosity, and branching patterns, are characteristic of certain eye diseases, which eye care professionals can monitor or diagnose \cite{fraz2012blood}. Manual measurements by ophthalmologists of these attributes are time-consuming, error-prone, and require specialist training. If automatic segmentation of retinal images using deep learning becomes more widely adopted, it will reduce the time needed for monitoring or diagnosis.

Image segmentation is the process of dividing an image into multiple segments or regions to extract relevant information for further analysis. There are various image segmentation techniques, including unsupervised methods, mathematical morphology, model-based methods, vessel tracking methods, supervised methods, traditional machine learning, deep learning methods, and combined methods.

One of the earliest works on image segmentation is traced back to Chaudhuri et al \cite{chaudhuri1989detection} who proposed the detection of blood vessels in retinal images using two-dimensional matched filters. They showed that the optimal filter, commonly known as the matched filter, maximizes the output signal-to-noise ratio. Applying this theory to 2-D images, they showed that the filters needed to be rotated at all possible angles, the corresponding outputs were to be compared, and only the maximum response for each pixel was to be retained. The result of applying these filters to retinal images produced better results than what was obtainable at the time.

Vessel tracking is another method used for image segmentation. Retinal blood vessel detection and tracking using Gaussian and Kalman filters was done by Chutatpe et al \cite{chutatape1998retinal}. They considered a blood vessel to be the linking of many small vessel segments and began the tracking process from the optic disc, which is the origin of all blood vessels. Although this method was able to detect more vessel networks in ocular fundus photographs, there were many drawbacks, one of which was that it does not provide segmentation in cases where the blood vessel fades away in the middle part and emerges again in its extended direction.

Mathematical morphology is another method used for image segmentation. Aswini et al \cite{aswini2018retinal} proposed the segmentation of retinal vessels using the top-hat morphological approach for the detection of diabetic retinopathy. They achieved an average accuracy of 95.12\% and 94.35\% on the HAGIS and HRD datasets, respectively, though other metrics such as precision and sensitivity were not impressive.

In medical imaging, the segmentation of retinal blood vessels is an important task for the diagnosis and treatment of eye diseases. Several methods have been proposed for this task, ranging from traditional machine learning techniques to deep learning methods. In this regard, this article summarizes some of the methods proposed for retinal blood vessel segmentation and their performance metrics.

The thresholding method, which is one of the earliest methods, applies a fixed threshold to the image's pixel intensity to segment blood vessels. Another method, the model-based method, is based on the Mumford-Shah (MS) model and skeletonization method, where the MS model finds a smooth function that is similar to the input image and minimizes it. Additionally, the combined method applies traditional machine learning techniques such as the random forest and convolutional neural network to classify if pixels correspond to real blood vessels or not.

Traditional probabilistic methods have been used for both optic disc and retinal blood vessel segmentation \cite{Kaba:his2013,Kaba:hiss2014}. Other statistical methods such as Markov Random Fields \cite{Salazar:his2012,Elizabeth2019,Eltayef:ida2017,Eltayef:cbms2017} have also been applied to this problem. 

Extensive studies on Graph Cut methods for retinal image analysis have been reported in \cite{Salazar:cimi2011,Salazar:his2012,Salazar:jbhi2014,Salazar:icarcv2010,Salazar:jaiscr2012} and \cite{Kaba:oe2015}. The Level-Set Method has also been successfully applied to many segmentation and medical imaging problems \cite{wang2020blood,Wang:jms2015,Wang:icig2015,Wang:jbhi2017,Wang:icig2015_2,dodo2020simultaneous,Dodo:jms2019,Dodo:best2019,Dodo:cbms2017,Dodo:cbms2019,Dodo:access2019,Dodo:bioimaging2018,Dodo:bioimaging2019,Dodo:bioimaging2018_2}. 

Deep learning methods have shown significant improvements in the segmentation of retinal blood vessels. The U-Net architecture \cite{ronneberger2015u} is a popular deep learning method applied in the segmentation of retina blood vessels. Researchers have made modifications to the original U-Net architecture by reducing two levels of pooling operations and introducing the Squeeze and Excitation block. Additionally, the multi-level convolutional neural network with different scales of input and four convolutional layers has been proposed, achieving an AUC value of 0.9782 \cite{guo2018automatic}.

Karaali et al \cite{karaali2022dr} proposed a supervised method called DR-VNet for retina vessel segmentation based on Convolutional Neural Network (CNN) with the aim of overcoming the problem of poor sensitivity rate while keeping other accuracy metrics at a high level. Their architecture consists of two cascaded sub-networks – a Backbone Residual Dense network and a Fine-tune Tail Network. They proposed a composite loss function that combines the traditional binary cross-entropy loss function and the dice loss, which is commonly used in medical imaging-related tasks.

Recently, variations of the U-Net \cite{ndipenoch2022simultaneous,ndipenoch2023retinal} and nnUNet \cite{mcconnell2022integrating} have been reported on this problem with improved results.

Overall, the performance of these methods varies in terms of accuracy, precision, sensitivity, and specificity. It is important to note that the development of more advanced and accurate methods for retinal blood vessel segmentation could significantly improve the diagnosis and treatment of eye diseases.

\section{Methods}

Four architectures are investigated in this project. They are UNet,  DR-VNet, UNet-ResNet and UNet-VGG.  
 
\subsection{The UNet}
This first architecture to be used  to perform the experiment is the simplest form of the U-Net developed by Ronneberger et al (2015). The network has 23  layers,  a  contracting  and  expansive  path.  The  contractive  path  has  two 3x3  unpadded  convolutions  followed  by  an  activation  layer,  a  2x2  max-pooling  operation  and  a  stride  of  2.  In  the  expansive  path,  the  number  of feature  channels  are  doubled,  concatenation  with  output  of  the  feature channels and two 3x3 convolution operations followed by the activation layer. 

\subsection{The DR-VNet}

The Dense Residual VNet, proposed by Karaali et al \cite{karaali2022dr}, is the second architecture employed in this experiment. It comprises of two consecutive sub-networks - the Backbone Residual Dense network and the Fine-tune Tail network.

The Backbone Residual DenseNet follows the U-shaped structure of U-Net, but with modifications. Instead of the original convolutional layers, it uses two blocks: the Residual Dense-net block and the Residual Squeeze and Excitation block. Additionally, it has four up-sampling and down-sampling blocks, compared to U-Net's three. The Fine-tune Tail network, on the other hand, is a shallow network consisting of three Residual DenseNets and Residual Squeeze and Excitation blocks. The Backbone Residual Network processes the images first, and the output is fine-tuned in the Fine-tune Tail network.

Each layer in the network architecture includes a Residual DenseNet (RDN) and a Residual Squeeze and Excitation block (RSE). The RDN is responsible for feature extraction using ResNet and DenseNet techniques, while the RSE modifies the weights of each channel to highlight the contribution of channels with more relevant features. It employs max-pooling layers with a kernel size of 2x2 and a stride of 2. The final output block has a single 1x1 convolution operation that generates an initial estimate of the vessel map.

In the Fine-tune Tail network, the first layer uses RDN and RSE techniques on the input and output of the Backbone Residual DenseNet sub-network. The output of this layer is concatenated and passed through a two-layer neural block consisting of RDN and RSE. The final map is determined by thresholding.

\subsection{Unet-ResNet and Unet-VGG}

Two additional models, namely Unet-ResNet34 and Unet-VGG19 \cite{yakubovskiy2018segmentation}, were used in this project. These models employ transfer learning, which is facilitated by the Segmentation Models library in Python for image segmentation using the Keras (Tensorflow) framework. This library contains four model architectures and 25 backbones for each architecture with pre-trained weights. For this experiment, the U-net model was used with ResNet34 and VGG19 backbones. This approach is considered to be an easier but efficient way of performing semantic segmentation, utilizing various pre-configured models and backbones. Backbones are named after classification models such as VGG, ResNet, SENet, DenseNet, Inception, MobileNet, and EfficientNet, all of which were trained on the 2012 ILSVRC ImageNet dataset.

\section{Experiments}

\subsection{Datasets}

Four publicly available datasets were combined manually to form the dataset of 88 training images and 45 testing images to perform the experiment. They are:

\begin{enumerate}
\item Digital Retinal Images for Vessel Extraction (DRIVE) \\
\url{https://drive.grand-challenge.org/}): This dataset, obtained from a diabetic retinopathy screening program in the Netherlands, contains 20 training images, 20 testing images, 2 sets of training ground truth labels and the corresponding test labels.

\item Structured Analysis of the Retina (STARE) \\
\url{https://cecas.clemson.edu/~ahoover/stare/})\\
 This dataset was obtained in 1975 under a clinical project in California, USA. It contains 20 images for training and 20 testing images.

\item The Child Heart and Health Study in England dataset (CHASE\_DB) \\
\url{https://blogs.kingston.ac.uk/retinal/chasedb1/}) \\
It consists of 28 retinal images of the left and right eye of multi-ethnic children divided into.

\item High Resolution Fundus (HRF) \\
\url{https://www5.cs.fau.de/research/data/fundus-images/}) \\
This contributed 37 training images and 8 testing images.
\end{enumerate}

\subsection{Data Pre-Processing}

The following steps were taken during the data pre-processing phase:

\begin{itemize}
\item \textbf{Image Formats}: The dataset to be used comprises of images of different formats and sizes, for example .tif, .gif, .jpg, .JPG. To enable learning efficiency, a function was used to read images of different formats and sizes.
\item \textbf{Detail Enhancement}: The Contrast Limited Adaptive Histogram Equalization (CLAHE) algorithm is used in this task to enhance the local contrast of images.
\item \textbf{Grayscale Conversion}: The fundus images have three channels - red (R), green (G) and blue (B). Theoretically, it’s been found that it’s better to use the green channel for analysis as it shows less noise and high contrast between the blood vessels and the background.
\item \textbf{Data Augmentation}: This is a technique of applying various methods on neural network training images to synthetically increase the number and diversity of training data (Hernandez-Garcia and Konig, 2018). Data augmentation will be implemented on the combined dataset. Using the Python Albumentations library, techniques implemented are horizontal flip, vertical flip and geometric rotation. The total number of training images increased from 88 to 264 after data augmentation was implemented.
\end{itemize}

\subsection{Evaluation Matrics}

The quantitative evaluation metrics which will be utilised are explained below as it relates to image segmentation:  

a)  Accuracy: In this task the equivalent accuracy metric used is the pixel 
accuracy.  It  reports  the  percentage  of  pixels  in  the  image  which  are correctly classified. A true positive (TP) represents a correctly predicted pixel classified as a blood vessel compared to the label, a true negative (TN)  represents  a  correctly  predicted  pixel  classified  as  a  non-blood vessel. The false positive (FP) and false negative (FN) are pixels which are  classified  as  blood  vessels  which  turn  out  to  be  false  and  pixels which aren’t classified as blood vessels but indeed turn out to be blood vessel pixels respectively.

b)  Precision:  This  metric  measures  the  correctness  of  the  predicted 
positive pixels relative to the ground truth.

c)  Recall:  This  measures  the  completeness  of  the  positive  pixel 
predictions relative to the ground truth. It reveals the total number of 
positive  predictions  that  was  done  by  the  model  compared  to  what’s 
available in the ground truth. 

d)  Intersection  over  Union:  This  is  also  called  the  Jaccard  index.  It quantifies  the  percentage  overlap  between  the  ground  truth  and  the prediction output by measuring the number of pixels common between 
the target and prediction masks divided by the total number of pixels 
present across both masks

e)  F1-Score: The F1-score is a metric used to compare the performance of 
multiple  classifiers  –  in  this  case,  the  architectures.  It  combines  the 
Precision and Recall by taking their harmonic mean. 

\subsection{Results and Discussions}

\begin{figure}
\begin{center}
\includegraphics[width=\textwidth]{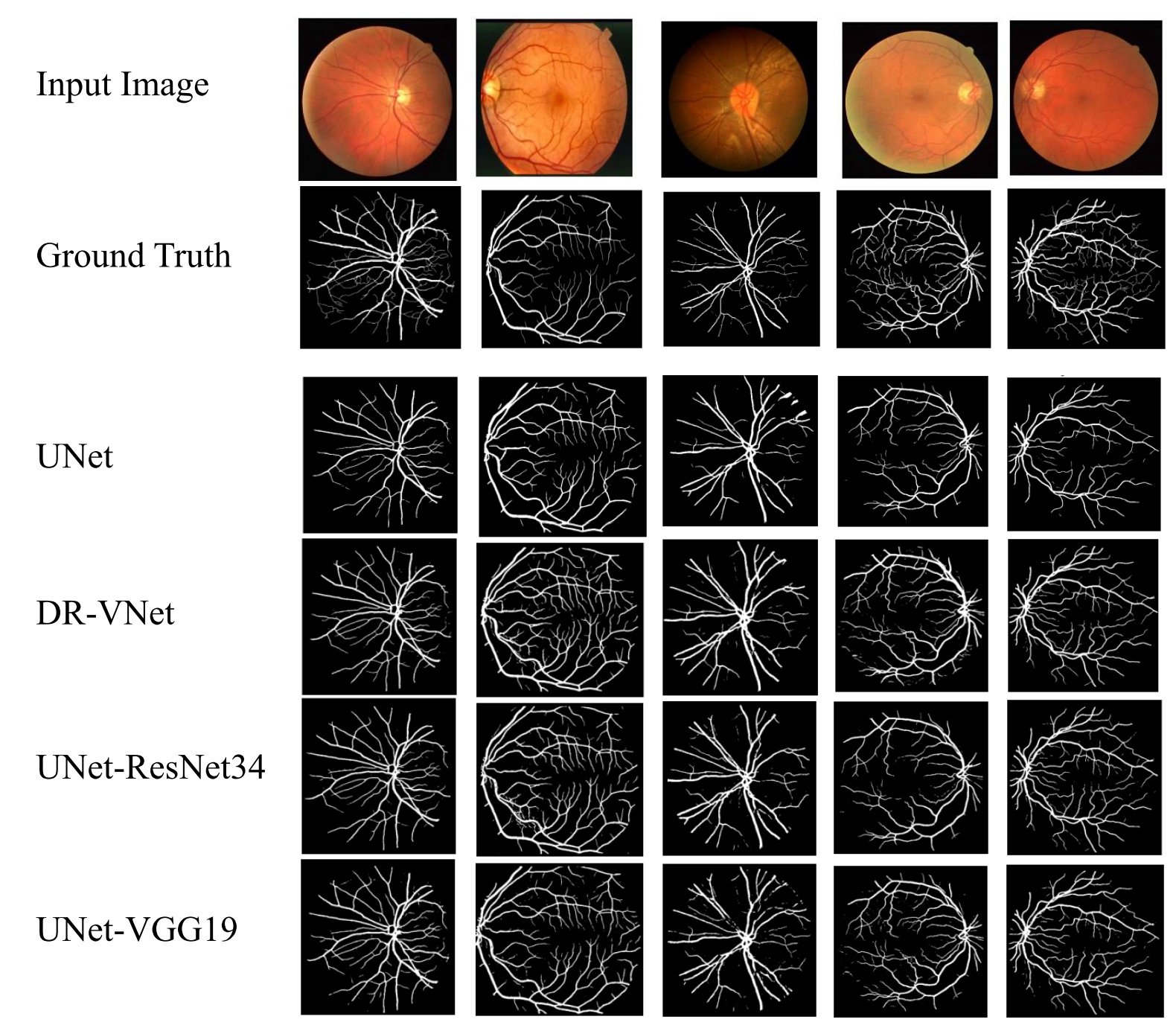}
\caption
{Sample results of blood vessel segmentation. From top to bottom are:
the original images, ground truth,  segmentation output from the four methods of UNet, DR-VNet, UNet-ResNet34 and UNet-VGG19.}
\label{fig:results}
\end{center}
\end{figure}

Experiments from  the four neural network architectures discussed in the previous section were carried out  on a manually combined dataset taken from four publicly available retina segmentation datasets. The results measured by the chosen quantitative metrics are listed in Table~\ref{tab:performance}. Samples of segmentation results from the four methods are shown in Figure~\ref{fig:results}.

\begin{table}[htbp]
    \centering
    \caption{Comparison of performance metrics for different methods}
    \label{tab:performance}
    \begin{tabular}{lccccc}
        \hline
        Method & Accuracy & Precision & Recall & F1-Score & Mean IoU \\
        \hline
        UNet & 0.9941 & 0.9243 & 0.8785 & 0.8921 & 0.8865 \\
        DR-VNet & 0.9602 & 0.7679 & 0.7170 & 0.7397 & 0.5882 \\
        UNet-ResNet & 0.9524 & 0.8236 & 0.5129 & 0.6221 & 0.4581 \\
        UNet-VGG19 & 0.9552 & 0.7967 & 0.5882 & 0.6645 & 0.5041 \\
        \hline
    \end{tabular}
\end{table}

\section{Conclusions}

There has been a disparity between the quantitative performance of the four 
architectures in predicting the blood vessels and the qualitative results of the segmented images.  
 
While the evaluation metrics showed that the conventional UNet architecture 
was  the  better  performing  model,  the  qualitative  observation  of  the 
segmented  images  showed  otherwise.  After  an  analysis  of  the  five  images from the test dataset, the Dense Residual V-Net (DR-VNet) architecture and the  UNET-VGG19  architecture  consistently  predicted  better  retinal  blood vessel  segmentations  when  thin  and  tiny  vessels  were  the  goal.  It  was observed that continuity of more blood vessels were observed with the DR-VNet model.  The  segmented  images  predicted  using  UNET-VGG19  were  observed  to  be bolder than the blood vessels in the ground truth.  

\bibliographystyle{abbrv}
\bibliography{bib2023}


\end{document}